\documentclass[preprint,12pt]{elsarticle}

\usepackage{amsmath,amsthm,amsfonts}
\usepackage{amssymb}
\usepackage{graphicx}
\usepackage{color}
\newtheorem{teo}{Theorem}[section]

\newtheorem{defn}{Definition}[section]
\newtheorem{alg}{Algorithm}[section]

\newcommand{\R}{\mathbb{R}}				% \R   = numeri reali
\DeclareMathOperator{\Sym}{Sym}			% \Sym = algebra simmetrica

\newcommand{\bfx}{\boldsymbol{x}}	
\newcommand{\bfalpha}{\boldsymbol{\alpha}}	

\journal{International Journal of Nonlinear Mechanics}

\begin{document}

\begin{frontmatter}
\title{A Ginzburg-Landau model for the expansion of a dodecahedral viral capsid}
\date{}

\author[dmy]{E. Zappa}
\ead{ez537@york.ac.uk}
\author[ycc]{G. Indelicato}
\ead{giuliana.indelicato@york.ac.uk}

\author[dmt]{A. Albano}
\ead{alberto.albano@unito.it}
\author[dmt]{P. Cermelli}
\ead{paolo.cermelli@unito.it}

\address[dmy]{Department of Mathematics, University of York, UK}
\address[ycc]{York Centre for Complex Systems Analysis, Department of Mathematics, University of York, UK}
\address[dmt]{Dipartimento di Matematica, Universit\`{a} di Torino, Italy}

\begin{abstract}
We propose a Ginzburg-Landau model for the expansion of a dodecahedral viral capsid during infection or maturation. The capsid is described as a dodecahedron whose faces, meant to model rigid capsomers, are free to move independent of each other, and has therefore twelve degrees of freedom. We assume that the energy of the system is a function of the twelve variables with icosahedral symmetry. Using techniques of the theory of invariants, we expand the energy as the sum of invariant polynomials up to fourth order, and classify its minima in dependence of the coefficients of the Ginzburg-Landau expansion. Possible conformational changes of the capsid correspond to symmetry breaking of the equilibrium closed form. The results suggest that the only generic transition from the closed state leads to icosahedral expanded form. Our approach does not allow to study the expansion pathway, which is likely to be non-icosahedral.

\end{abstract}

\end{frontmatter}

\section{Introduction}

Most viruses are made of a protein shell, the capsid, built of identical protein units, that encapsidates and hence protects the nucleic acid (RNA or DNA). Although viruses exhibit a wide diversity of shape a large number of them display icosahedral symmetry and this is irrespective of the number and the chemical composition of the protein subunits, the capsomers, that constitute the capsid. The basic principles to account for the icosahedral arrangement of the protein in a capsid were outlined by Caspar and Klug in the quasi-equivalence theory \cite{Caspar1962}, and this continues to be a fundamental framework in virology.

During their life-cycle viruses undergo structural transitions. These phenomena can be triggered by a change of the environment, such as a variation of temperature or pH. The occurence of a transitions induces a radial expansion of the capsid and a rearrangement of the capsomers  and consequently the opening of pores on the capsid, so that the genetic material is exposed and eventually released in the host cell \cite{robinson,Sherman,Speir,tuthill}.
 
In this work we focus on icosahedral viruses whose expansion can be modelled through the independent motion of twelve pentagonal blocks. These can be viral particles  whose capsomers are pentamers (group of five proteins) or more complex viruses whose protein subunit are arranged to form pentagonal blocks \cite{tuthill}. 

In general, the energy of the capsid should  account both for the cohesive forces between capsomers and for an internal pressure that  tends to expand the capsid.  However, instead of making specific choices, we adopt here the Ginzburg-Landau approach, and assume that the energy of the capsid is a function of the twelve independent degrees of freedom of the capsomers, and its explicit expression is given in terms of  invariant polynomials of the icosahedral group. The coefficients of the expansion are the basic control parameters of our model and, in turn, they should  be related to the environment of the capsid. We truncate the expansion at the fourth degree, which yields a sufficiently rich energy landscape to account for the basic physics of the model.  

We relax the symmetry conditions and classify the minima of the energy according to their symmetry, in dependence of the parameters. The appearance of new minima corresponds to conformational changes of the capsid, with  symmetry possibly lower than icosahedral. We restrict to minima of the energy corresponding to one of the three maximal  subgroups of the icosahedral group: the tetrahedral group, and the dihedral groups of order 6 and 10.

The analysis shows  that only four of the nine expansion parameters are relevant in this model: the parameter space is subdivided into regions, in which minima have a given symmetry. The main result of our work is that the only generic transition from a closed configuration with icosahedral symmetry turns out to be to an icosahedral expanded state. Minima with lower symmetry are not accessible from the closed reference state because they involve the cooperative change of more than one control parameters. 

Hence,  our study shows that a phenomenological model based on the Ginzburg-Landau expansion is able to describe one of the main features of the expansion of a viral capsid: even though the transition pathway is unlikely to be icosahedral (cf. \cite{indelicato,cermelli}),  the final equilibrium state is still icosahedral  (cf. \cite{cermelli}).

Our work complements other approaches to the study of conformational changes of viral capsids, either based on coarse-graining and phenomenological interaction potentials between the capsomers and relying on domain-decomposition techniques  \cite{micheletti,cermelli},
or on the geometrical description of the fine features of the capsid via libraries of point sets with icosahedral symmetry \cite{Keef2006,Keef3,KT, indelicato}, normal mode analysis of the atomic  ensemble of the capsid \cite{tama}  or, finally, using the continuum theory of thin shells \cite{guerin-bruinsma,bruinsma0}.

\section{ Basics of the Ginzburg-Landau approach}\label{landau}
Let us consider a system described by state variables $(x_1, \ldots, x_n)=\bfx   \in \mathbb{R}^n$ and a \emph{symmetry group} $G$ acting on $\mathbb{R}^n$. The action of $G$ provides a representation $\rho : G \longrightarrow GL(\mathbb{R},n)$. We associate to the system a \emph{free energy}
\begin{equation}
\begin{split}
E :  \;  \mathbb{R}^n\times A  & \longrightarrow \mathbb{R} \\
(\bfx, \boldsymbol{\alpha}  ) &\longmapsto E(\bfx,\boldsymbol{\alpha}  ),
\end{split}
\label{energy1}
\end{equation}
where $A \subseteq \mathbb{R}^m$ is an open subset,  $\boldsymbol{\alpha} \in A$ are parameters affecting the system (i.e., temperature, pH, etc.), and $E\in C^2(\mathbb{R}^n\times A)$.
 We require the energy to be invariant with respect to $\rho$, i.e.
\begin{equation}
E( \rho(g)\bfx,\boldsymbol{\alpha}) = E( \bfx ,\boldsymbol{\alpha} ) \qquad \forall g \in G, \quad  \forall \bfx   \in \mathbb{R}^n \quad \forall \boldsymbol{\alpha} \in A.
\label{energy2}
\end{equation}
The minima of $E(\cdot,\boldsymbol{\alpha})$ with respect to $\bfx  $ correspond to the \emph{stable phases} of the system. Let us introduce the following definition
\begin{defn}
Let $\rho: G \longrightarrow GL( \mathbb{R}^n)$ be a representation of a finite group $G$. The \emph{isotropy subgroup} of $\bfx   \in  \mathbb{R}^n$ is
\begin{equation*}
\Sigma_{\bfx  } = \{g \in G : \rho(g)\bfx   = \bfx   \}.
\end{equation*}
\end{defn} 
Minima of $E(\cdot,\boldsymbol{\alpha})$ can be classified according to their isotropy group.  In particular, if  $\bfx_0$  is a minimum of $E(\cdot,\boldsymbol{\alpha})$ such that 
\begin{equation*}
\Sigma_{\bfx_0}=G,
\end{equation*}
we say that the system is in a high-symmetry phase. In general,  we are interested in studying the local minima of the energy as $\boldsymbol{\alpha}$ varies. The system undergoes a phase transition when the number and the symmetry of the minima changes as $\boldsymbol{\alpha}$ varies \cite{rif1}. 

Notice that the level sets of $E(\cdot,\boldsymbol{\alpha})$ are invariant under $\rho$. Therefore, if there exists a minimum $\bfx_0$ such that
\begin{equation*}
\Sigma_{\bfx_0}=H,\qquad H< G,
\end{equation*}
all points of the orbit $\rho(G)\bfx_0$ also are minima. If $G$ is finite, the orbit $\rho(G)\bfx_0$ is also finite and has $|G/H|$ elements.  Hence, low symmetry phases occur in different variants, while  high symmetry phases are in general unique. 

Since we are interested in icosahedral viruses 
we consider the icosahedral group $\mathcal{I}$ which consists of all the rotations which leave a regular icosahedron invariant. It has order 60, and it is generated by the elements $g_2$ (twofold rotation) and $g_5$ (fivefold rotation) such that $(g_2)^2 = (g_5)^5 = (g_2 g_5)^3 = e$, the identity element. It is isomorphic to the alternating group of order 5, $A_5$. 

Moreover $\mathcal{I}$ acts naturally on the twelve faces of a dodecahedron by permutation. In this way, we obtain a function $\sigma : \mathcal{I} \longrightarrow S_{12}$,
where $S_{12}$ is the symmetric group of order 12, such that
\begin{equation*}
\begin{split}
&\sigma(g_{2}) = 
(1,6)(2,5)(3,9)(4,10)(7,12)(8,11),
\\
&
\sigma(g_{5}) = (1,2,3,4,5)(7,8,9,10,11).
\end{split}
\end{equation*}
It is easy to see that $\sigma$ is an homomorphism, since
\begin{equation*}
\sigma(g_2)^2 = \sigma(g_5)^5 = (\sigma(g_2) \sigma(g_5))^3=\text{id}_{S_{12}}.
\end{equation*}
$\sigma$ is called a \emph{permutation representation} of $\mathcal{I}$, being an homomorphism between a group and a permutation group. The symmetric group $S_{12}$ acts naturally on $\mathbb{R}^{12}$ by permuting the indices of a vector.
This action induces a representation
$\rho : \mathcal{I} \longrightarrow GL(\mathbb{R},12)$ 
such that $\rho(g_2)^2 = \rho(g_5)^5 = (\rho(g_2)\rho(g_5))^3 = I_{12}$, with $I_{12}$ the identity matrix of order 12.

\section{Formulation of the model}\label{model}
We now formulate a model for the expansion of the capsid in the framework of the Ginzburg-Landau theory. 

\begin{figure}[!h]
\centering
\includegraphics[scale=0.3]{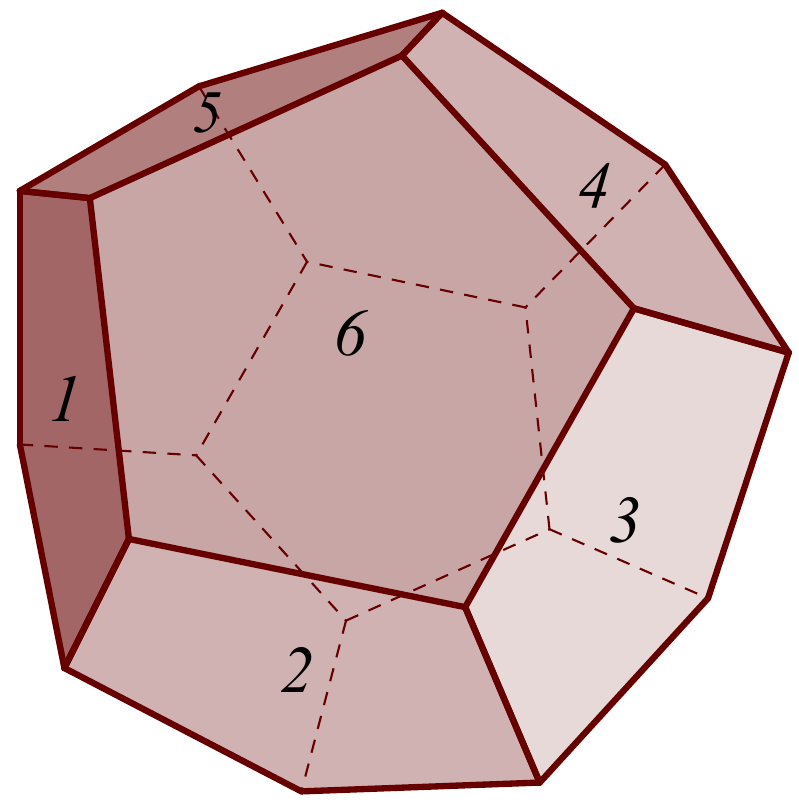}
\caption{A dodecahedron: each pentagonal face is labelled by a number from 1 to 12.  The face opposite to face $i$ is labelled by $i+6$, for $i=1,\ldots,6$.
}
\label{dodecaedro}
\end{figure}

\begin{figure}[!h]
\centering
\includegraphics[scale=1]{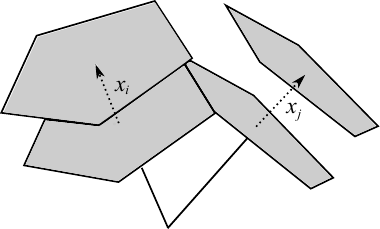}
\caption{Basic variables of the model.
}
\label{traslaz}
\end{figure}

We focus here on  viruses  with capsid with icosahedral symmetry, whose expansion can be described via the independent motion of twelve pentagonal blocks. Therefore, before expansion, the capsid can be modelled as a regular dodecahedron (Figure \ref{dodecaedro}). We associate to each capsomer, labeled by $i=1,\ldots,12$, a translation parameter $x_i$ along the axis of the pentagonal face, with $x_i \in\R$ (cf. Figure \ref{traslaz}).  The translation variables $x_i$ have meaning only when non negative, but we allow them to assume negative values in order to simplify the analysis of the minima of the energy. However, we will only accept as physically meaningful the minima whose components are all non negative.

The translation parameters $x_i, i=1, \ldots, 12$, represent the state variables of the system, i.e., $\bfx   = (x_1, \ldots ,x_{12})$.
Note that $\mathbf{0}=(0,\ldots,0)$ represents the closed configuration of the capsid, which has by definition full icosahedral symmetry, i.e., its isotropy subgroup is the whole group $\mathcal{I}$. 

We require the energy $E$ to be a ${C}^{2}$ function $E : \mathbb{R}^{12}\times A \rightarrow \mathbb{R}$, with $A\subset\R^m$, such that:
\begin{itemize}
\item[(i)]  it is invariant with respect to the representation $\rho$ of $\mathcal{I}$ on $\R^{12}$, i.e.,
\begin{equation*}
E(\rho(g) \bfx ,\bfalpha )=E(\bfx,\bfalpha  ) \qquad \forall g \in \mathcal{I}, \quad \forall \bfalpha \in A, \quad \forall \bfx   \in \mathbb{R}^{12}.
\end{equation*}
\item[(ii)] $\mathbf{0}=(0, \ldots, 0)$ is a critical point of $E$, i.e.
\begin{equation*}
\nabla_{\bfx} E(\mathbf{0},\bfalpha) = \mathbf{0} \qquad \forall \bfalpha \in A.
\end{equation*}
\item[(iii)] \begin{equation*}
\lim_{|\bfx  |\rightarrow +\infty}E(\bfx,\bfalpha)=+ \infty\qquad \forall \bfalpha \in A.
\end{equation*}
\end{itemize}
As a first approach we also require the energy to be  polynomial in the variables $x_i$. This assumption relies on the fact that, if the energy is sufficently smooth, it can be expanded as a Taylor polynomial in a neighbourhood of $\mathbf{0}$, which still has the property of invariance with respect to $\rho$. 

Our goals are:
\begin{itemize}
\item to find explicit expressions of the energy $E$ as the sum of icosahedrally invariant polynomials;

\item to find the minima  of $E$ in dependence of the parameters $\bfalpha$. When $\mathbf{0}$, which is always an extremum due to hypoyhesis (ii), is a minimum, the capsid is in the closed configuration. We interpret the appearance of new minima of the energy as conformational changes of the capsid associated to its expansion, driven by variations  of the parameters $\bfalpha$, that in turn are related to the environment of the capsid.

\end{itemize}

\section{Explicit form of the energy}
In this section we describe a method to find an explicit form of the energy. As we pointed out in the previous section, we assume it to be a  polynomial invariant under the action of $\mathcal{I}$. Therefore, we first study the ring of invariant polynomials and its structure. Complete proofs for the assertions below can be found, for example, in \cite{rif4} or \cite{rif11}.

Let $V$ be a vector space over a field~$K$ of characteristic zero (in what follows, we will be interested mainly in $K = \R$) with $\dim V = n$. We fix a basis $\mathcal{A}$ of $V$ and let $\{x_1$, \dots, $x_n\}$ be the dual basis of $V^*$.

Let $G$ be a finite group and $\rho : G \longrightarrow GL(V)$ a representation of $G$. The action of $G$ on~$V$ induces the dual action on~$V^*$ and hence an action on the symmetric algebra~$\Sym(V^*) \cong K[x_1, \dots, x_n]$. We denote by $K[\bfx  ]^G = K[x_1, \ldots, x_n]^G$ the subring of polynomials invariant under the action of $G$.
$K[\bfx  ]$ is a graded ring, i.e., 
\begin{equation*}
K[\bfx  ] = \bigoplus_{k=0}^{+\infty}K[\bfx  ]_k
\end{equation*}
where $K[\bfx  ]_k$ is the vector space of the homogeneous polynomials of degree~$k$.
Since the action of $G$ preserves the grading, $K[\bfx  ]^G$ is a graded subring and we can write
\begin{equation*}
K[\bfx  ]^G = \bigoplus_{k=0}^{+\infty}K[\bfx  ]^G_k
\end{equation*}
where $K[\bfx  ]^G_k=K[x_1, \ldots, x_n]^G_k$ is the set of all the invariant polynomials of degree $k$.

One way to find invariant polynomials is to take the average over the orbits of the action. More precisely, let us introduce the so-called Reynolds operator
\begin{equation*}
R_G : K[\bfx  ] \longrightarrow K[\bfx  ]^G
\end{equation*}
defined by
\begin{equation*}
R_G(f)(\bfx  ) = \frac{1}{|G|}\sum_{g \in G}(g \cdot f)(\bfx  ).
\end{equation*}

It is clear that $R_G$ is a $K$-linear map that is the identity on the subring~$K[\bfx  ]^G$ (in particular, it is surjective). Hilbert showed, as a consequence of his famous Basis Theorem, that

\begin{teo}[Hilbert] The invariant ring $K[\bfx  ]^G$ of a finite group~$G$ is finitely generated.
\end{teo}

Later E. Nother proved that the degree of the generators is bounded above by the order of the group:

\begin{teo}[E. Noether] Let $|G| $ be the order of the group~$G$. Then $K[\bfx  ]^G$ is generated, as a $K$-algebra, by (finitely many) polynomials of degree less or equal than~$|G|$.
\end{teo}

Hence there exist $p_1, \ldots, p_t \in K[\bfx  ]^G$ such that $K[\bfx  ]^G = K[p_1, \ldots, p_t]$, i.e., every polynomial in $K[\bfx  ]^G$ can be written as a polynomial in the $p_i$ and $\deg p_i \le |G|$ for~$i = 1, \dots, t$. Moreover, we can find the generators by applying the Reynolds operator to a basis of the space of polynomials of degree less or equal than~$|G|$, for example to the monomials.

This gives the following

\begin{alg}\label{27}
In order to find a vector space basis of $K[\bfx  ]^G_k$, apply the Reynolds operator to all monomials in $K[\bfx  ]_k$. This yields a generating set of $K[\bfx  ]_k^G$ as a vector space. By linear algebra, a basis can be extracted from it.

To find a generating set for the full invariant ring, apply the previous step for $k = 1, 2, \dots, |G|$.
\end{alg}

This algorithm has been performed successfully using the computer algebra software SINGULAR \cite{rif12}. However, in the case where the order of $G$ is large and the representation of $G$ has a high degree, computations can be difficult, due to the high dimension of $K[\bfx  ]_k$. One way to determine some invariant polynomials with quite efficient computations is to use the irreducible representations of the group, as we are now going to discuss.  The result is by no means the most general invariant polynomial, but has the advantage of being amenable to further computations.

Let $\rho : G \longrightarrow GL(V)$ be a representation of $V$, and let $\mathcal{A}$ be a basis of $V$, with $\text{dim}V = n$. Using Maschke's theorem, we know that there exist  irreducible representations (defined over the algebraic closure of~$K$) $\rho_i : G \longrightarrow GL(V_i)$, $i=1 \ldots r$, with $V = V_1 \oplus \ldots \oplus V_r$, such that $\rho = \rho_1 \oplus \ldots \oplus \rho_r$. Given a basis $\mathcal{B}_i$ of $V_i$, we have
\begin{equation*}
[\rho(g)]_{\mathcal{B}}= \left(\begin{array}{ccc} [\rho_1(g)]_{\mathcal{B}_1} & \ldots & \mathbf{0} \\
\vdots & \ddots & \vdots \\
\mathbf{0} & \ldots & [\rho_r(g)]_{\mathcal{B}_r}
\end{array}\right)
\end{equation*} 
where $\mathcal{B}=\mathcal{B}_1 \cup \ldots \cup \mathcal{B}_r$ and $[\rho(g)]_{\mathcal{B}}$ is the matrix of $\rho(g)$ in the basis $\mathcal{B}$. Moreover, if $P$ is the matrix representing the change of basis from $\mathcal{A}$ to $\mathcal{B}$, we have 
\begin{equation*}
[\rho(g)]_{\mathcal{B}}=P^{-1}[\rho(g)]_{\mathcal{A}}P \qquad \forall g \in G.
\end{equation*}
We denote by $\bfx'=(\eta_1^{(1)},\ldots, \eta_{\text{dim}V_1}^{(1)}, \eta_1^{(2)}, \ldots ,\eta_{\text{dim}V_2}^{(2)}, \ldots, \eta_1^{(r)}, \ldots \eta_{\text{dim}V_r}^{(r)})$ and $\bfx  =(x_1, \ldots, x_n)$ the coordinates of a vector in $\mathbb{R}^n$ in the bases $\mathcal{B}$ and $\mathcal{A}$, respectively, so that $\bfx   = P\bfx  '$. \\
We consider the two rings of invariants $K[\bfx  ]^G$ and $K[\bfx  ']^G$. More precisely, $p \in K[\bfx  ]^G$ if
\begin{equation*}
p([\rho(g)]_{\mathcal{A}} \bfx  ) = p(\bfx  ) \qquad \forall g \in G
\end{equation*}
while $f \in K[\bfx  ']^G$ if 
\begin{equation*}
f([\rho(g)]_{\mathcal{B}} \bfx  ') = f(\bfx  ') \qquad \forall g \in G.
\end{equation*}
The two rings are clearly isomorphic, and an isomorphism is the following
\begin{align*}
\varphi : K[\bfx']^G & \; \longrightarrow K[\bfx  ]^G \\
f(\bfx') & \; \longmapsto \bar{f}(\bfx  ) = f(P^{-1}\bfx  ).
\end{align*}
In this way, it is possible to find polynomials of the ring $K[\bfx  ]^G$ by first finding polynomials of the ring $K[\bfx  ']^G$ and then using the function $\varphi$.

 Let $\mathcal{B}_i=\{\boldsymbol{v}_j^{(i)}\}_j$, $j = 1 \ldots \text{dim}V_i$. We have, for every vector $\bfx  ' \in \mathbb{R}^n$, since $V = V_1 \oplus \ldots \oplus V_r$
\begin{equation*}
\bfx  ' = \bfx  '_1 + \ldots + \bfx  '_r
\end{equation*}
with 
\begin{equation*}
\bfx'_i = \sum_{j = 1}^{\text{dim}V_i}\eta_j^{(i)}\boldsymbol{v}_j^{(i)}\equiv (\eta_1^{(i)}, \ldots, \eta_{\text{dim}V_i}^{(i)}).
\end{equation*}
Since $[\rho]_{\mathcal{B}} = [\rho_1]_{\mathcal{B}_1} \oplus \ldots \oplus [\rho_r]_{\mathcal{B}_r}$, invariance of $f$ can be rewritten as
\begin{equation*}
f([\rho_1]_{\mathcal{B}_1}\bfx  '_1+\ldots +[\rho_r]_{\mathcal{B}_r} \bfx  '_r) = f(\bfx  '_1 + \ldots + \bfx  '_r).
\end{equation*}
Therefore, we have
\begin{equation*}
\bigoplus_{i=1}^r K[\bfx  '_i]^G \subseteq K[\bfx  ']^G.
\end{equation*}
In other words, if $p^{(i)}(\bfx  '_i)=p^{(i)}(\eta_1^{(i)}, \ldots ,\eta_{\text{dim}V_i}^{(i)}) \in K[\bfx  '_i]^G$, i.e.
\begin{equation*}
p^{(i)}([\rho_i(g)]_{\mathcal{B}_i} \bfx  '_i) = p^{(i)}(\bfx  '_i) \qquad \forall g \in G,
\end{equation*}
then the polynomial
\begin{equation*}
\sum_{i=1}^r p^{(i)}(\bfx  '_i)
\end{equation*}
is an invariant polynomial in $K[\bfx  ']^G$. This is a sufficient condition for the invariance which is useful for computations.

Let us consider $K[\bfx  '_i]^G_k$, $k>0$. Using Algorithm \ref{27}, it is possible to find a basis $\{p^{(i)}_{kj}\}_j$ of $K[\bfx  '_i]^G_k$ and write
\begin{equation*}
p^{(i)}_k(\bfx  '_i)=\sum_j c_j p^{(i)}_{kj}(\bfx  '_i), \qquad c_j \in K.
\end{equation*}
Since $\text{dim}V_i < \text{dim}V$ the computations are easier (sometimes in a considerable way). We then consider the polynomial
\begin{equation*}
f(\bfx  ') = \sum_{k=1}^d \left( \sum_{i=1}^r p^{(i)}_k(\bfx  '_i) \right)
\end{equation*}
and finally using the function $\varphi$ we find $\varphi(f)=\bar{f}(\bfx  ) = f(P^{-1}\bfx  ) \in K[\bfx  ]^G$. \\
We organize the above results in the following algorithm

\begin{alg}
Let $V$ be a vector space with $\text{\rm dim} V = n$, and $\mathcal{A}$ a basis of $V$. Let $\rho: G \longrightarrow GL(V)$ be a representation of a finite group $G$, and $K[x_1, \ldots, x_n]^G = K[\bfx  ]^G$ the ring of the polynomials invariant under the action of $G$. Perform the following steps:
\begin{enumerate}
\item find the decomposition $\rho_1 \oplus \ldots \oplus \rho_r$ using the character table of the group $G$ and the formula (see \cite{rif6})
\begin{equation}\label{10}
\chi_{\rho} = \chi_{\rho_1}+ \ldots + \chi_{\rho_r};
\end{equation}
\item find explicitly the decomposition $V = V_1 \oplus \ldots \oplus V_r$ finding a basis $\mathcal{B}_i$ for each $V_i$ using the projection operator (see \cite{rif6}) $P_i : V \longrightarrow V_i$ defined by 
\begin{equation*}
P_i = \sum_{g \in G} \overline{\chi_{\rho_i}(g)}  \rho(g);
\end{equation*} 
\item find $[\rho_i]_{\mathcal{B}_i}$, $i=1, \ldots, r$ and, letting $\mathcal{B} = \mathcal{B}_1 \cup \ldots \cup \mathcal{B}_r$, find the matrix $P$ of the change of basis from $\mathcal{A}$ to $\mathcal{B}$ such that $[\rho]_{\mathcal{B}} = P^{-1}[\rho]_{\mathcal{A}} P$;
\item having fixed the degree $d>0$ and denoting by $\bfx  '$ the coordinates of a vector in the basis $\mathcal{B}$, find the polynomials $p^{(i)}_{kj}(\bfx'_i)$ forming a basis of $K[\bfx'_i]^G_k,$  using Algorithm \ref{27}, for $ i=1, \ldots r$ and $k=1, \ldots, d$;
\item write the polynomial 
\begin{equation*}
f(\bfx  ') = \sum_{k=1}^d \left( \sum_{i=1}^r \left( \sum_{j=1}^{t_{i,k}} c_j^{(i)}p^{(i)}_{kj}(\bfx  '_i) \right) \right)
\end{equation*}
where $t_{i,k} = \text{\rm dim}K[\bfx  '_i]^G_k$ and $c_j^{(i)} \in K$;
\item perform the change of variables and find $\bar{f}(\bfx  ) = f(P^{-1}\bfx  ) \in K[\bfx  ]^G$.
\end{enumerate}

\end{alg}
Notice that the above algorithm does not produce ``mixed'' invariants, that are the product of polynomials invariant under different irreducible representations of the group. We neglect such invariants in the following development.

We now apply this algorithm to our model. 
In our case, $G$ is the icosahedral group $\mathcal{I}$, and   $\rho : \mathcal{I} \longrightarrow GL(\mathbb{R},12)$ is the representation introduced in Section \ref{landau}. Since $\mathcal{I}$ is isomorphic to the alternating group $A_5$, its character table is the following
\begin{center}
\begin{tabular}{l|c c c c c}
Irrep & $\mathcal C(e)$ & $\mathcal C(g_{5})$ & $\mathcal C(g_{5}^{2})$ & $\mathcal C(g_{2})$ & $\mathcal C(g_2g_5)$ \\
\hline
$\rho_1$ & 1 & 1 & 1 & 1 & 1 \\
$\rho_2$ & 3 & $\tau$ & 1-$\tau$ & -1 & 0 \\
$\rho_3$ & 3 & 1-$\tau$ & $\tau$ & -1 & 0 \\
$\rho_4$ & 4 & -1 & -1 & 0 & 1 \\
$\rho_5$ & 5 & 0 & 0 & 1 & -1 \\
\end{tabular}
\end{center}
where $\tau = \frac{1+\sqrt{5}}{2}$, and $\mathcal{C}$ stands for conjugacy class. We note that, up to isomorphisms, there are 5 irreducible representations of $\mathcal{I}$. We also note that since all characters are real, the direct sum decomposition is defined over~$\R$.\\
In our case, we have $\chi_{\rho}(g_2)=0$ and $\chi_{\rho}(g_5)=2$.  We note that $1+\tau+(1-\tau)+0=2$, and $1-1-1+1=0$, so that
\begin{equation*}
\chi_{\rho} = \chi_{\rho_1}+\chi_{\rho_2}+\chi_{\rho_3}+\chi_{\rho_5}.
\end{equation*} 
Using formula \eqref{10}, we find that $\rho= \rho_1 \oplus \rho_2 \oplus \rho_3 \oplus \rho_5$. Therefore, we have the decomposition
\begin{equation*}
\mathbb{R}^{12} = V_1 \oplus V_2 \oplus V_3 \oplus V_5,
\end{equation*} 
where $\rho_i : \mathcal{I} \longrightarrow GL(V_i)$, and $\text{dim}V_2=\text{dim}V_3=3$, $\text{dim}V_1 = 1$, and $\text{dim}V_5=5$. It is possible to compute the projection operators $P_i$, $i = 1,2,3,5$, and, as a consequence, a basis $\mathcal{B}_i = \{\boldsymbol{v}_j^{(i)}\}_j$ for each $V_i$, considering first the set $\{P_i\mathbf{e}_j\}_{j=1}^{12}$, with $\mathcal{A}=\{\boldsymbol{e}_j\}_j$ the canonical basis of $\mathbb{R}^{12}$, and extracting a basis from it. We can then find $[\rho_i]_{\mathcal{B}_i}$ and the matrix $P$, whose columns are the vectors $\boldsymbol{v}_j^{(i)}$, $i = 1,2,3,5$, $ j = 1 \ldots \text{dim}V_i$, representing the change of basis from $\mathcal{A}$ to $\mathcal{B}= \mathcal{B}_1 \cup \mathcal{B}_2 \cup \mathcal{B}_3 \cup \mathcal{B}_5$. The results are
\begin{itemize}
\item {\bf Representation $\rho_1$} \\
\begin{displaymath}
[\rho_1(g_2)]_{\mathcal{B}_1}=[\rho_1(g_5)]_{\mathcal{B}_1}=(1)
\end{displaymath}
\item {\bf Representation $\rho_2$} 
\begin{equation*}
[\rho_2(g_2)]_{\mathcal{B}_2}=\left(
\begin{array}{ccc}
0 & -1 & 0 \\
-1 & 0 & 0 \\
0 & 0 & -1
\end{array}
\right),
\quad
[\rho_2(g_5)]_{\mathcal{B}_2}=\left(
\begin{array}{ccc}
1 & -\tau & -1 \\
0 & -1 &  -\tau\\
0 & \tau & \tau
\end{array}
\right),
\end{equation*}
\item {\bf Representation $\rho_3$} 
\begin{equation*}
[\rho_3(g_2)]_{\mathcal{B}_3}=\left(
\begin{array}{ccc}
1-\tau & 1-\tau & 0 \\
-1 & \tau-1 & 0\\
1-\tau & 1 & -1
\end{array}
\right)
,\quad
[\rho_3(g_5)]_{\mathcal{B}_3}=\left(
\begin{array}{ccc}
0 & 0 & 1 \\
1 & 0 & \tau-1 \\
0 & 1 & 1-\tau
\end{array} 
\right),
\end{equation*}
\item {\bf Representation $\rho_5$}
\begin{equation*}
[\rho_5(g_2)]_{\mathcal{B}_5}=\left(
\begin{array}{ccccc}
1 & 0 & 0 & 0 & 0 \\
0 & 0 & 0 & 0 & 1 \\
0 & 0 & 0 & 1 & 0 \\
0 & 0 & 1 & 0 & 0 \\
0 & 1 & 0 & 0 & 0 
\end{array}
\right),
\quad
[\rho_5(g_5)]_{\mathcal{B}_5}=\left(
\begin{array}{ccccc}
0 & 0 & 0 & 0 & -1 \\
1 & 0 & 0 & 0 & -1 \\
0 & 0 & 1 & 0 & -1 \\
0 & 1 & 0 & 0 & -1 \\
0 & 0 & 0 & 1 & -1 
\end{array}
\right),
\end{equation*}
\item  {\bf Projection matrix}
\begin{equation*}
{\tiny P= \left(
\begin{array}{cccccccccccc}
1 & -1 & \sqrt{5} & 1 & \sqrt{5} & -1 & 1 & 1 & 1 & 1 & -5 & 1 \\
1 & -1 & 1 & 1 & -1 & \sqrt{5} & -1 & 1 & 1 & 1 & 1 & -5 \\
1 & -1 & -1 & -1 & 1 & -1 & \sqrt{5} & 1 & 1 & 1 & 1 & 1 \\
1 & -1 & -1 & -\sqrt{5} & 1 & 1 & -1 & -5 & 1 & 1 & 1 & 1 \\
1 & -1 & 1 & -1 & -1 & 1 & 1 & 1 & -5 & 1 & 1 & 1 \\ 
1 & -\sqrt{5} & 1 & -1 & -1 & -1 & -1 & 1 & 1 & -5 & 1 & 1 \\
1 & 1 & -\sqrt{5} & -1 & -\sqrt{5} & 1 & -1 & 1 & 1 & 1 & -5 & 1 \\ 
1 & 1 & -1 & -1 & 1 & -\sqrt{5} & 1 & 1 & 1 & 1 & 1 & -5 \\  
1 & 1 & 1 & 1 & -1 & 1 & -\sqrt{5} & 1 & 1 & 1 & 1 & 1 \\ 
1 & 1 & 1 & \sqrt{5} & -1 & -1 & 1 & -5 & 1 & 1 & 1 & 1 \\
1 & 1 & -1 & 1 & 1 & -1 & -1 & 1 & -5 & 1 & 1 & 1 \\
1 & \sqrt{5} & -1 & 1 & 1 & 1 & 1 & 1 & 1 & -5 & 1 & 1 \\   
\end{array}
\right).}
\end{equation*}
\end{itemize}
The following holds
\begin{equation*}
[\rho(g)]_{\mathcal{B}}=P^{-1}[\rho(g)]_{\mathcal{A}}P=\left(\begin{array}{cccc} [\rho_1(g)]_{\mathcal{B}_1} & \ldots & \ldots & \mathbf{0} \\
\vdots & [\rho_2(g)]_{\mathcal{B}_2} & \vdots & \vdots \\
 \vdots &  \vdots & [\rho_3(g)]_{\mathcal{B}_3} & \vdots \\
\mathbf{0} & \ldots & \ldots & [\rho_5(g)]_{\mathcal{B}_5}
\end{array}\right)
\end{equation*} 
We are now able to find the invariant polynomials of degree $k$, $p^{(i)}_{kj}$, forming a basis of $\mathbb{R}[x_1, \ldots ,x_{\text{dim}V_i}]^{\mathcal{I}}_k$, using Algorithm \ref{27}. For the representation $\rho_2$ and $\rho_3$, which are $3$-dimensional, we write the polynomials in the variables $x,y,z$, and we perform the calculations for $k=2,4,6$. There are no invariant polynomials of degree $1,3,5$. For the representation $\rho_5$, which is $5$-dimensional, we use the variables $x,y,z,t,u$. In this case we find invariant polynomials of degree $2,3,4$.\\

\begin{itemize}
\item {\bf{Representation $\rho_1$}} 
\\
\begin{equation*}
p^{(1)}_{k1} = x^k \qquad k = 1, 2, \ldots
\end{equation*}
\item \noindent {\bf{Representation $\rho_2$}} 
\\
 \noindent $k=2$
\begin{equation*}
p_{21}^{(2)}(x,y,z) =  x^2  +y^2+z^2+\frac{2\sqrt{5}}{5}\left(xz+yz-xy\right) \\
\end{equation*}
$k=4$
\begin{align*}
p_{41}^{(2)}(x,y,z) = & \left(p_{21}^{(2)}\right)^2
\end{align*}
\item {\bf{Representation $\rho_3$}} 
\\

\noindent $k=2$
\begin{equation*}
p_{21}^{(3)}(x,y,z) =  x^2  +y^2+z^2+\frac{2\sqrt{5}}{5}\left(xz-yz-xy\right) 
\end{equation*}
$k=4$
\begin{align*}
p_{41}^{(3)}(x,y,z) = \; &  \left(p_{21}^{(3)}\right)^2
\end{align*}
\item {\bf{Representation $\rho_5$}}
\\
\noindent $k=2$
\begin{equation*}
\begin{split}
p_{21}^{(5)}(x,y,z,t,u) &= x^2+y^2+z^2+t^2+u^2
\\
-\frac{2}{5}(xy+& xz+yz+xt+yt+zt+xu+yu+zu+tu)
\end{split}
\end{equation*}
$k=3$
\begin{align*}
p_{31}^{(5)}(x,y,z,t,u) =\; & xyz-xyt-xzt+yzt+xyu
-xzu-yzu\\&+xtu-ytu+ztu \\
p_{32}^{(5)}(x,y,z,t,u) =\; & x^3+y^3+z^3+ t^3 +u^3+\frac{6}{5}(xyt+xzt+xzu+yzu+ytu)\\&-\frac{3}{5}(x^2y+xy^2+x^2z 
 +y^2z+xz^2+yz^2+x^2t+y^2t\\&+z^2t+xt^2+ yt^2+zt^2+x^2u+y^2u \\
&+z^2u+t^2u+xu^2+yu^2+zu^2+tu^2) 
\end{align*}
$k=4$
\begin{align*}
p_{41}^{(5)}(x,y,z,t,u) = \; & x^2y^2+x^2z^2+y^2z^2+x^2t^2+z^2t^2+x^2u^2+t^2u^2\\&+z^2u^2+y^2u^2+y^2t^2 \\
&+xyzt+xyzu+xytu+xztu+yztu\\&-\frac{1}{2}(x^2yz+xy^2z+xyz^2+x^2yt 
+xy^2t+x^2zt\\&+y^2zt+xz^2t +yz^2t+xyt^2+xzt^2+yzt^2+x^2yu \\
& +xy^2u+x^2zu+y^2zu+xz^2u+yz^2u+x^2tu+y^2tu\\
&+z^2tu+xt^2u +yt^2u+zt^2u +xyu^2+xzu^2+yzu^2\\
&+xtu^2+ytu^2+ztu^2) 
\end{align*}

\begin{align*}p_{42}^{(5)}(x,y,z,t,u) = \; & x^4+y^4+z^4+t^4+u^4-\frac{4}{5}(x^3y+xy^3+x^3z+y^3z\\
&+x^3t+xz^3+yz^3 +y^3t+z^3t+xt^3+yt^3+zt^3\\
&+x^3u+y^3u+z^3u+t^3u+xu^3+yu^3 
 +zu^3+tu^3)\\
&+\frac{3}{5}(x^2yz+xy^2z+xyz^2 +x^2yt+xy^2t+x^2zt+y^2zt\\
 & +xz^2t+yz^2t+xyt^2 +xzt^2+yzt^2+x^2yu+xy^2u\\
&+x^2zu +y^2zu  +xz^2u+yz^2u+x^2tu+y^2tu+z^2tu\\
&+xt^2u+yt^2u +z^2tu+xt^2u  
  +yt^2u + zt^2u+ xyu^2\\
&+ xzu^2+ yzu^2+ xtu^2+ ytu^2+ztu^2)\\
&-\frac{6}{5}(xyzt+xyzu+xytu+xztu+yztu)
\end{align*}

\end{itemize}

Let us now consider the basis $\mathcal{B}=\{\boldsymbol{v}_j^{(i)}\}_{i,j}$. Let us denote by
\begin{equation*}
 \bfx  ' = \sum_{i,j}\eta_j^{(i)}\boldsymbol{v}_j^{(i)}\equiv (\eta_1^{(1)}, \eta_1^{(2)},\eta_2^{(2)},\eta_3^{(2)},\eta_1^{(3)},\eta_2^{(3)},\eta_3^{(3)},\eta_1^{(5)},\eta_2^{(5)},\eta_3^{(5)}, \eta_4^{(5)},\eta_5^{(5)})
\end{equation*}
a vector in $\mathbb{R}^{12}$ written in the basis $\mathcal{B}$. Clearly, we have
\begin{equation*}
\bfx  ' = \bfx  '_1+\bfx  '_2+\bfx  '_3+\bfx  '_5
\end{equation*}
where
\begin{equation*}
\bfx  '_i = \sum_{j=1}^{\text{dim}V_i}\eta_j^{(i)}\boldsymbol{v}_j^{(i)}\equiv (\eta_1^{(i)}, \ldots ,\eta_{\text{dim}V_i}^{(i)}) \in V_i \qquad i=1,2,3,5.
\end{equation*}
We can write an invariant polynomial $F(\bfx  ') \in \mathbb{R}[\bfx  ']^{\mathcal{I}}$ of degree $d>0$ as a sum 
\begin{equation}\label{12}
F^{(d)}(\bfx  ')=\sum_{k=1}^d\left(\sum_{i=1,2,3,5}\left(\sum_jc_jp^{(i)}_{kj}(\bfx  '_i)\right)\right).
\end{equation}
The last step is to perform the change of coordinates from $\mathcal{B}$ to $\mathcal{A}$. If we denote by $\bfx  =(x_1, \ldots ,x_n)$ the coordinates of a vector in $\mathbb{R}^{12}$ in the canonical basis $\mathcal{A}$, we know that $\bfx  =P\bfx  '$.
Performing this change of variables in \eqref{12}, we find 
\begin{equation*}
E^{(d)}(\bfx  )=F^{(d)}(P^{-1}\bfx  )=\sum_{k=1}^d\left(\sum_{i=1,2,3,5}\left(\sum_jc_j^{(i)}p^{(i)}_{kj}(\eta^{(i)}_1(\bfx  ), \ldots \eta_{\text{dim}V_i}^{(i)}(\bfx  ))\right)\right)
\end{equation*}
where $c_j^{(i)} \in \mathbb{R}$. $E^{(d)}(\bfx  )$ is an invariant polynomial in $\mathbb{R}[\bfx  ]^{\mathcal{I}}$ of degree $d>0$.

We then choose $d=4$ and, for semplicity, we discard the three polynomials of third degree $p_{31}^{(1)}, p_{31}^{(5)}$ and $p_{32}^{(5)}$.  Further,  in order for $\mathbf{0}$ to be an extremum of the energy function, 
the linear terms $p_{11}^{(1)}(\bfx  ) = \sum_{i=1}^{12}x_i$ must be absent. 

The output of the procedure is therefore a polynomial invariant energy of degree 4, which will be the focus of our subsequent analysis: 
\begin{equation}\label{14}
\begin{split}
E(\bfx ,\bfalpha) =& a p_{21}^{(1)} + b p_{21}^{(2)} + c p_{21}^{(3)} + d p_{21}^{(5)} 
\\
&+ c_1 p_{41}^{(1)} + c_2 p_{41}^{(2)} + c_3 p_{41}^{(3)} + c_4 p_{41}^{(5)} + c_5 p_{42}^{(5)}
\end{split}
\end{equation}
where
\begin{equation}\label{parameters}
\bfalpha = (a,b, c,d,c_1,c_2 ,c_3 ,c_4,c_5)\in\R^{9}
\end{equation}
and we assume that $c_i>0$, $i=1,\ldots,5$ so that (iii) in Section \ref{model} holds.

\section{Bifurcation analysis}
In this section we characterize the minima of the energy in dependence of the parameters $\bfalpha$. The stationary points of the energy function are solutions of the system
\begin{equation}
\nabla_{\bfx} E(\bfx,\bfalpha  )=\mathbf{0}.
\label{15}
\end{equation}
We look for solutions of (\ref{15}) with given nontrivial isotropy subgroup.   The subgroups of $\mathcal{I}$ are listed in Table \ref{sottogruppi}, and the permutation  representations of their generators are listed in Table  \ref{permrep2}.
\begin{table}[ht]
\begin{center}
\begin{tabular}{l c c c}
Subgroup & Generators & Relations & Order \\
\hline
${T}$ & $g_2, g_{3d}$ & $g_2^2=g_{3d}^3=(g_2g_{3d})^3=e$ & 12 \\
${D}_{10}$ & $g_{2d},g_{5d}$ & $g_{2d}^2=g_{5d}^5=(g_{5d}g_{2d})^2=e$ & 10   \\
${D}_6$ & $g_{2d},g_3$ & $g_{2d}^2=g_3^3=(g_3g_{2d})^2=e$ & 6 \\
$\mathbb{Z}_5$ & $g_5$ & $g_5^5=e$ & 5  \\
${D}_{4}$ & $g_{2d},g_2$ & $g_{2d}^2=g_2^2=(g_2g_{2d})^2=e$ & 4  \\
$\mathbb{Z}_3$ & $g_3$ & $g_3^3=e$ & 3 \\
$\mathbb{Z}_2$ & $g_2$ & $g_2^2=e$ & 2 \\
\hline
\end{tabular}
\end{center}
\caption{The subgroups of the icosahedral group. We have used the notation  $T$ for the tetrahedral group, $D_{2n}$ for the dihedral group of order $2n$, and $\mathbb{Z}_n$ for  the cyclic group of order $n$.}
\label{sottogruppi}
\end{table}
For a subgroup $H$ of $\mathcal{I}$,  consider the linear subspace of $\mathbb{R}^{12}$ whose elements are the vectors fixed by  $H$:
\begin{equation}\label{16}
\text{Fix}(H) = \{ \bfx   \in \mathbb{R}^{12} : \rho(g) \bfx   = \bfx   \quad \forall g \in H \}.
\end{equation}
Clearly,  $\bfx   \in \text{Fix}(H)$ if and only if  $H<\Sigma_{\bfx  } $.   Therefore, in order to find minima of the energy with minimal symmetry $H$, it is enough to solve (\ref{15}) in $\text{Fix}(H)$, i.e.,
letting  $ \boldsymbol{\gamma} : \mathbb{R}^k \longrightarrow \text{Fix}(H)\subset \mathbb{R}^{12}$  a parametrization of $\text{Fix}(H)$, with $k=\dim(\text{Fix}(H))$, to  solve
\begin{equation}
(\nabla_{\bfx}E \circ \boldsymbol{\gamma})(\boldsymbol{t}) = \mathbf{0}, \qquad \boldsymbol{t}\in\R^k.
\label{17}
\end{equation}
This system  has $k$ independent equations only, since, omitting the dependence on $\bfalpha$ for simplicity, the following holds
\begin{equation*}
(\nabla_{\bfx} E \circ \boldsymbol{\gamma} )(\boldsymbol{t}) \in \text{Fix}(H), \qquad \forall \boldsymbol{t} \in \mathbb{R}^k.
\end{equation*}
To see this, notice first that,  since $E$ is invariant with respect to  $\rho$, then $\nabla_{\bfx} E$ is {equivariant}, i.e.
\begin{equation*}
\nabla_{\bfx} E(\rho(g)\bfx  ) = \rho(g)\nabla_{\bfx} E(\bfx  ) \qquad \forall \bfx   \in \mathbb{R}^{12},
\end{equation*}
so that, for $\boldsymbol{t} \in \mathbb{R}^k$, 
\begin{equation*}
\begin{split}
\rho(g) (\nabla_{\bfx} E \circ \boldsymbol{\gamma})(\boldsymbol{t}) &= \rho(g) \nabla_{\bfx} E(\boldsymbol{\gamma}(\boldsymbol{t})) = \nabla_{\bfx} E(\rho(g)\boldsymbol{\gamma}(\xi)) 
\\
&= \nabla_{\bfx} E(\boldsymbol{\gamma}(\boldsymbol{t}))  = (\nabla_{\bfx} E \circ \boldsymbol{\gamma})(\boldsymbol{t}),
\end{split}
\end{equation*} 
which is the assertion to be proved.

Let $\sigma_1 =\sigma(g_1), \ldots,  \sigma_r = \sigma(g_r)$ be the permutation representations of the generators  $g_1 \ldots g_r$ of $H$.  Each $\sigma_i \in S_{12}$ can be decomposed into disjoint cycles, i.e. $\sigma_i = (n_1 n_2 \ldots n_s)(m_1 m_2 \ldots m_r)$ $ \ldots (p_1 p_2 \ldots p_t)$. Let  $\bfx   = (x_1,\ldots,x_{12}) \in \mathbb{R}^{12}$. Recalling that $\sigma_i \cdot \bfx   = (x_{\sigma_i(1)}, \ldots x_{\sigma_i(12)})$, if
\begin{equation}\label{19}
\left \{ \begin{array}{l}
x_{n_1}= x_{n_2}=\ldots x_{n_s} \\
x_{m_1}= x_{m_2}=\ldots x_{m_r}\\
 \vdots \\
 x_{p_1}= x_{p_2}=\ldots x_{p_t} 
\end{array} \right.
\end{equation}
then $\sigma_i \cdot \bfx   = \bfx  $. Iterating this argument for all generators of the subgroup $H$ we can find a parametrization of  $\text{Fix}(H)$.

\begin{table}[ht]
\begin{center}
\begin{tabular}{lcc}
\hline
$\sigma(g_2) $&$= $&$ (1 , 6)(2 , 5)(3 , 9)(4 , 10)(7 , 12)(8 , 11)$,
\\
$\sigma(g_{2d}) $&$= $&$
(1,12)(2,8)(3,4)(5,11)(6,7)(9,10)$,
\\
$\sigma(g_{3})$&$ = $&$
(1,2,6)(3,5,10)(4,9,11)(7,8,12)$,
\\
$\sigma(g_{3d})$&$ = $&$
(1,10,2)(3,5,12)(4,8,7)(6,9,11)$,
\\
$\sigma(g_5) $&$=$&$  (1 , 2 , 3 , 4 , 5)(7 , 8 , 9, 10, 11)$,
\\
$\sigma(g_{5d}) $&$= $&$
(1,10,11,3,6)(4,5,9,12,7)$.
\\
\hline
\end{tabular}
\end{center}
\caption{Permutation representations and disjoint cycle decompositions of the generators of the subgroups of the icosahedral group.}
\label{permrep2}
\end{table}

Table \ref{FIX} lists  $\text{Fix}(H)$ for all subgroups of  the icosahedral group. We only describe here the computations of $\text{Fix}(D_{10})$. In this case, the generators are $g_{5d}$ and $g_{2d}$, and  \eqref{19} becomes
\begin{equation*}
\left \{ \begin{array}{l}
x_1 = x_{10} = x_{11} = x_3 = x_6 \\
x_4 = x_5 = x_9 = x_{12} = x_7 
\end{array} \right.,
\qquad
\left \{ \begin{array}{l}
x_1 = x_{12} \\
x_2 = x_8 \\
x_3 = x_4 \\
x_5 = x_{11} \\
x_6 = x_7 \\
x_9 = x_{10}
\end{array} \right.,
\end{equation*}
which combined yield
\begin{equation*}
\left \{ \begin{array}{l}
x_1 = x_3 = x_4 = x_5 = x_6 = x_7 = x_9 = x_{10} = x_{11} = x_{12}=x, \\
x_2 = x_8 = y,
\end{array} \right.
\end{equation*}
which is the result in Table \ref{FIX}.

\begin{table}[ht]
\begin{center}
\begin{tabular}{lcl}
\hline
$\text{Fix}(T )$&$ = $&$\{ \bfx   = (x,x,x,x,x,x,x,x,x,x,x,x) : x \in \mathbb{R} \}$,
\\
$\text{Fix}(D_{10}) $&$=$&$ \{ \bfx   = (x,y,x,x,x,x,x,y,x,x,x,x) : x,y \in \mathbb{R} \}$,
\\
$\text{Fix}(D_6)$&$=$&$ \{ \bfx   = (x,x,y,y,y,x,x,x,y,y,y,x) : x,y \in \mathbb{R} \}$,
\\
$\text{Fix}(\mathbb{Z}_5)$&$ = $&$\{ \bfx   = (x,x,x,x,x,y,z,z,z,z,z,t) : x,y,z,t \in \mathbb{R} \}$,
\\
$\text{Fix}(D_4) $&$=$&$ \{ \bfx   = (x,y,z,z,y,x,x,y,z,z,y,x) : x,y,z \in \mathbb{R} \}$,
\\
$\text{Fix}(\mathbb{Z}_3) $&$=$&$ \{ \bfx   = (x,x,y,z,y,x,t,t,z,y,z,t) : x,y,z,t \in \mathbb{R} \}$,
\\
$\text{Fix}(\mathbb{Z}_2) $&$= $&$\{ \bfx   = (x,y,z,t,y,x,u,w,z,t,w,u) : x,y,z,t,u,w \in \mathbb{R} \}$.
\\
\hline
\end{tabular}
\end{center}
\caption{Fixed subspaces in $\R^{12}$ of the subgroups of the icosahedral group. 
Notice that vectors in $\text{Fix}(T )$ have full icosahedral symmetry.}
\label{FIX}
\end{table}
 
 \subsection{Stability of the closed capsid: conditions for the origin to be a minimum.}

Consider the energy \eqref{14}. Since there are no terms of first degree, $\mathbf{0}$ is always a stationary point of $E$ for any $\bfalpha$.

The hessian matrix  of $E$ evaluated at $\mathbf{0}$ has eigenvalues $\frac{1}{6}a$ (multiplicity 1), $\frac{1}{10}b$ (multiplicity 3), $\frac{1}{10}c$ (multiplicity 3), and $\frac{1}{30}d$ (multiplicity 5). Therefore,
$$
a,b,c,d>0
\quad
\Leftrightarrow
\quad
\mathbf{0} \text{ is a minimum}
$$
and the closed capsid is stable. 

\subsection{Symmetry--breaking stable solutions}

We restrict our analysis to the maximal subgroups of $\mathcal{I}$: the tetrahedral group $T$, and the dihedral groups $D_{10}$ and $D_{6}$, which correspond to minimal symmetry breaking.

 \subsubsection{Solutions with icosahedral symmetry $T $ or $\mathcal{I}$}
In this case the system (\ref{17}) reduces to
\begin{equation*}
\frac{1}{6}ax+\frac{1}{3}c_1x^3 = 0.
\end{equation*}
If $a>0$ it has only the trivial solution, while if $a<0$ (we always suppose $c_i>0$ for the positive definiteness of the energy) we find $x=\pm \sqrt{\frac{-a}{2c_1}}$. We are only interested in strictly positive solutions, corresponding to the expansion of the capsid, so that we are left with the stationary point 
\begin{equation}\label{icosahedral}
\bfx  _0 = (\xi, \ldots, \xi), \qquad \xi = \sqrt{-\frac{a}{2c_1}}.
\end{equation}
The eigenvalues of the Hessian matrix $H(\bfx_0)$ of $E$ computed at $\bfx_0$  differ from those of $H(\mathbf{0})$ only for the first one, which is now $-\frac{1}{3}a$. Therefore, when $a<0$, $\mathbf{0}$ loses its stability and a new minimum, $\bfx_0$, appears, which corresponds to a complete expansion of the capsid which retains  full icosahedral symmetry.   The value of the energy computed at the icosahedral solution is 
$$
E(\bfx_0,\alpha)=-\frac{a^2}{4c_1}.
$$

\subsubsection{Solutions with fivefold symmetry $D_{10}$}
In this case the system (\ref{17}) reduces to
\begin{equation*}
\left \{ \begin{aligned}
&\frac{a}{36}(5x+y)+\frac{d}{180}(x-y)+\frac{c_1}{648}(5x+y)^3+\frac{c_5}{3240}(x-y)^3 = 0,\\
&\frac{a}{36}(5x+y)+\frac{d}{36}(y-x)+\frac{c_1}{648}(5x+y)^3+\frac{c_5}{648}(y-x)^3 = 0,
\end{aligned} \right.
\end{equation*}
i.e.,
\begin{equation*}\label{24-1}
\left \{ \begin{aligned}
&(x-y)\left((x-y)^2+\frac{18 d }{c_5}\right) = 0\\
&(5x+y)\left((5x+y)^2+\frac{18 a }{c_1}\right)=0.
\end{aligned} \right.
\end{equation*}
Assuming as usual that $c_1,c_5>0$, for  $a\ge 0$ and $d\ge0$ the only solution is $x=y=0$. For
 $a > 0$ and  $d<0$ the solution  $5x+y=0$ is not acceptable because one of the variables would be negative. For  $a < 0$ e $d\ge0$ the solution has icosahedral symmetry, and we recover the solution (\ref{icosahedral}): 
 $$
 y=x=\sqrt{-\frac{a}{2 c_1}}.
 $$ 
 For  $a < 0$ and $d<0$ the solutions are 
\begin{equation}\label{49e}
\left \{ \begin{array}{l}
x=\sqrt{-\frac{a}{2c_1}}+\sqrt{-\frac{d}{2c_5}}\\
y=\sqrt{-\frac{a}{2c_1}}-5\sqrt{-\frac{d}{2c_5}}
\end{array} \right.,
\qquad
\left \{ \begin{array}{l}
x=\sqrt{-\frac{a}{2c_1}}-\sqrt{-\frac{d}{2c_5}}\\
y=\sqrt{-\frac{a}{2c_1}}+5\sqrt{-\frac{d}{2c_5}}.
\end{array} \right.
\end{equation}
Both components of the first  solution are positive for $a/c_1<25d/c_5<0$, and both components of the second solution are positive for   $a/c_1<d/c_5<0$. Denoting by $\bfx_1,\bfx_2\in\R^{12}$ 
the corresponding points, the eigenvalues of $H(\bfx  _1)$ are $-\frac{1}{15}d$, $-\frac{1}{3}a$,$\frac{1}{10}c$ (multiplicity 3), $\frac{1}{10}b$ ( multiplicity 3) and $\frac{d}{1440}(54-25c_4/c_5)$ (multiplicity 4). Therefore, if $c_4>\frac{54}{25}c_5$, and $b,c>0$, $\bfx  _1$ is a minimum. 
Moreover, $H(\bfx  _2)$ has the same eigenvalues as $H(\bfx  _1)$, and therefore if $c_4>\frac{54}{25}c_5$ and $b,c>0$, also $\bfx  _2$ is a minimum.

To summarize, with the assumptions 
\begin{equation*}
\frac{a}{c_1} \leq 25\frac{d}{c_5} <0,\quad 
b,c>0,\quad c_1 , c_5 >0,\quad 
c_4 > \frac{54}{25}c_5,
\end{equation*}
there are two stable phases $\bfx_1$ and $\bfx_2$  after the expansion, whose isotropy subgroup is $D_{10}$. The corresponding value of the energy is
\begin{equation*}
E(\bfx  _1) = E(\bfx  _2) = -\frac{c_5a^2+c_1d^2}{4c_1c_5}.
\end{equation*}
Instead, in the case where
\begin{equation*}
25\frac{d}{c_5} < \frac {a}{c_1} <\frac{d}{c_5}<0, \quad b,c>0,\quad c_1 ,c_5 >0,\quad 
c_4 > \frac{54}{25}c_5,
\end{equation*}
there is only one acceptable stable phase, corresponding to $\bfx  _2$. 

Notice that for each solution considered here all other points in the icosahedral orbit are also solutions with the same symmetry.  For instance, while  $\bfx_0$ has icosahedral symmetry and therefore it is a full orbit,  $\bfx_1$ and $\bfx_2$ have symmetry $D_{10}$, so that each corresponds to $60/10-1=5$ other equilibria with the same symmetry.

\subsubsection{Solutions with threefold symmetry $D_6$}
We proceed as in the previous case. The system (\ref{17})
reduces to
\begin{equation*}
\left \{ \begin{aligned}
&\frac{a}{12}(x+y)+\frac{d}{60}(x-y)+\frac{c_1}{24}(x+y)^3+\frac{c_4}{1296}(x-y)^3 = 0 \\
&\frac{a}{12}(x+y)-\frac{d}{60}(x-y)+\frac{c_1}{24}(x+y)^3-\frac{c_4}{1296}(x-y)^3 = 0, 
\end{aligned} \right.
\end{equation*}
i.e., 
\begin{equation*}
\left \{ \begin{aligned}
&(x+y)\left(\frac{2a}{c_1}+(x+y)^2\right) = 0 \\
&(x-y)\left(\frac{108d}{5c_4}+(x-y)^2\right) =0.
\end{aligned} \right.
\end{equation*}
Assuming that $c_1,c_4>0$, for  $a\ge 0$ and $d\ge0$ the only solution is $x=y=0$. For
 $a > 0$ and  $d<0$ the solution  $x+y=0$ is not acceptable because one of the variables would be negative. For  $a < 0$ e $d\ge0$ the solution has icosahedral symmetry, as in  (\ref{icosahedral}),
 $$
 y=x=\sqrt{-\frac{a}{2 c_1}}.
 $$ 
 For  $a < 0$ and $d<0$ the solutions are 
\begin{equation*}
\left \{ \begin{array}{l}
x_3 = \sqrt{-\frac{a}{2c_1}}+\frac{3}{5}\sqrt{-\frac{15d}{c_4}}\\
y_3 = \sqrt{-\frac{a}{2c_1}}-\frac{3}{5}\sqrt{-\frac{15d}{c_4}} 
\end{array} \right.,
\qquad
\left \{ \begin{array}{l}
x_4 = \sqrt{-\frac{a}{2c_1}}-\frac{3}{5}\sqrt{-\frac{15d}{c_4}} \\
y_4 = \sqrt{-\frac{a}{2c_1}}+\frac{3}{5}\sqrt{-\frac{15d}{c_4}}
\end{array} \right..
\end{equation*}
Both components of the first  solution are positive for $a/c_1 \leq \frac{54}{5}d/c_4$, and both components of the second solution are positive in the same range  $a/c_1 \leq \frac{54}{5}d/c_1$. Denoting by $\bfx_3,\bfx_4\in\R^{12}$ 
the corresponding points, the eigenvalues of $H(\bfx_3)$ are $-\frac{1}{15}d$, $-\frac{1}{3}a$, $\frac{1}{10}c$  (multiplicity 3), $\frac{1}{10}b$ (multiplicity 3), and 
$\frac1{1200}d(25-54\frac{c_5}{c_4})$ (multiplicity 4). Hence $\bfx  _3$ is a minimum if $c_5>\frac{25}{54}c_4$ and $b,c>0$

Finally, the hessian matrix $H(\bfx  _4)$ has the same eigenvalues as $H(\bfx  _3)$. Therefore, if $c_5>\frac{25}{54}c_4$ and $b,c>0$, also $\bfx  _4$ is a minimum.

We conclude that, with the assumptions 
\begin{equation*}
\frac{a}{c_1} \leq \frac{54}{5}\frac{d}{c_4} <0,\quad 
b,c>0,\quad c_1 , c_4 > 0,\quad 
c_5> \frac{25}{54}c_4,
\end{equation*}
there are two stable phases $\bfx_3$ and $\bfx_4$  after the expansion, whose isotropy subgroup is $D_6$.  The corresponding energy is
\begin{equation*}
E(\bfx  _3)=E(\bfx  _4) =-\frac{c_4a^2+\frac{54}{25}c_1d^2}{4c_1c_4}.
\end{equation*}

\section{Conclusions}

We have explored in this work  the possibility of using a Ginzburg-Landau approach to model conformational changes of viral capsids with icosahedral symmetry.  Using a simplified model of the capsid as a dodecahedron whose faces, intended to model rigid capsomers, can move independently, we have assumed that the energy of the capsid is 4th degree polynomial invariant under the icosahedral group. The explicit form of the polynomial has been determined using the irreducible representations of the icosahedral group.   The variables of the model are the amount of translation of each pentagonal face along its axis and, by consequence, we will only restrict to minima with nonnegative components.

\begin{figure}[!h]
\centering
\includegraphics[scale=0.9]{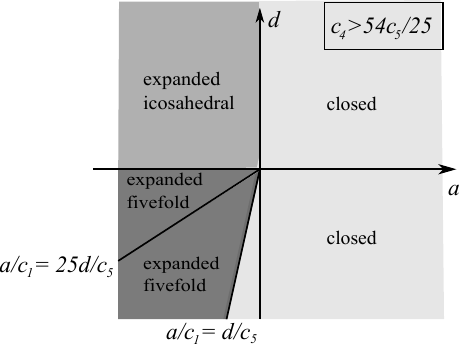}
\includegraphics[scale=0.9]{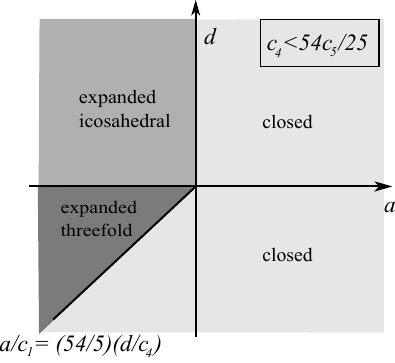}
\caption{Bifurcation diagrams for the minima of the energy for $b,c,c_1,c_2,c_3,c_4,c_5>0$.}
\label{bifurcations}
\end{figure}

In this context, the basic parameters of the model are 
the coefficients of the polynomials corresponding to each irreducible representations, and the conformational change can be viewed as the change of structure and symmetry of the minima of the energy as the parameters vary.  The phase diagram is depicted in Figure \ref{bifurcations}: when all parameters are positive, or when $d<0$ all other parameters being positive, the only minimum of the energy corresponds to the closed form of the capsid, but other minima appear, with non-icosahedral symmetry, in selected regions of the parameter space.

It is believed that conformational changes in viral capsids occur via non-icosahedral pathways, but that the final form after  maturation (of course when the capsid does not disassembly) is icosahedral. What is then the meaning of our non-icosahedral minima? We view the expansion process as a conformational change of the closed, stable form of the capsid triggered by variations of the chemical environment of the virion. The destabilization of the capsid occurs by changing some or all the parameters in the energy 
function, but, if the initial closed form corresponds to a point in parameter space in which all parameters are strictly positive, it is most unlikely that the minima cross special points of the boundary from the closed-form region to one of the low-symmetry regions. The most likely situation is that the minimum crosses a boundary, such as $a=0$,  and retains icosahedral symmetry.  

In other words, the conformational change leading to an icosahedral expanded state in viral capsids is preferred because it is generic, and therefore `easier': only one of the parametrs must change sign in order for an expanded icosahedrally symmetric state to appear.

\section*{Acknowledgements}
We thank Reidun Twarock and Giovanni Zanzotto for useful discussions on the topic of this work.
GI thanks the Leverhulme Trust for financial support via a Research Leadership Award. 
PC and GI also acknowledge the Italian PRIN 2009 project  ``Mathematics and Mechanics of Biological Systems and Soft Tissues".

%\cite{rif1}\cite{rif2}\cite{rif3}\cite{rif4}\cite{rif5}
%\cite{rif6}\cite{rif7}\cite{rif8}\cite{rif9}\cite{rif10}\cite{rif11}\cite{rif12}
%

\bibliographystyle{elsarticle-num}
%\nocite{*}
\bibliography{zcai}

\end{document}